%% file: MCX_QFT_arxiv.tex
\documentclass[sn-mathphys-num]{sn-jnl}

\usepackage{graphicx}%
\usepackage{multirow}%
\usepackage{amsmath,amssymb,amsfonts}%
\usepackage{amsthm}%
\usepackage{mathrsfs}%
\usepackage[title]{appendix}%
\usepackage[dvipsnames]{xcolor}%
\usepackage{textcomp}%
\usepackage{manyfoot}%
\usepackage{booktabs}%
\usepackage{algorithm}%
\usepackage{algorithmicx}%
\usepackage{algpseudocode}%
\usepackage{listings}%

\begin{document}

\title[Article Title]{Implementing multi-controlled X gates using the quantum Fourier transform}


\author*[1]{\fnm{Vladimir V.} \sur{Arsoski}}\email{vladimir.arsoski@etf.bg.ac.rs}

\affil*[1]{\orgdiv{The Department of Microelectronics and Technical Physics}, \orgname{School of Electrical Engineering - University of Belgrade}, \orgaddress{\street{Bulevar kralja Aleksandra 73}, \city{Belgrade}, \postcode{P.O. Box 35–54}, \country{Serbia}}}

\abstract{Quantum computing has the potential to solve many complex algorithms in the domains of optimization, arithmetics, structural search, financial risk analysis, machine learning, image processing, and others. Quantum circuits built to implement these algorithms usually require multi-controlled gates as fundamental building blocks, where the multi-controlled Toffoli stands out as the primary example. For implementation in quantum hardware, these gates should be decomposed into many elementary gates, which results in a large depth of the final quantum circuit. However, even moderately deep quantum circuits have low fidelity due to decoherence effects and, thus, may return an almost perfectly uniform distribution of the output results. This paper proposes a different approach for efficient cost multi-controlled gates implementation using the quantum Fourier transform. We show how the depth of the circuit can be significantly reduced using only a few ancilla qubits, making our approach viable for application to noisy intermediate-scale quantum computers. This quantum arithmetic-based approach can be efficiently used to implement many complex quantum gates.}

\keywords{Quantum computing, Quantum algorithms, Multi-controlled gates, Quantum Fourier transform, Ancilla qubits}


\pacs[MSC Classification]{03G12, 81P68}

\maketitle

\section{Introduction}\label{sec1}

Recent progress in the fabrication of quantum computers brings the accelerated development of quantum algorithms that can use the laws of quantum mechanics to solve some tasks faster than classical algorithms. It is well-known that the classes of problems
related to quantum chemistry and quantum physics are very challenging for classical computation \cite{Feynman1982}. When solved using quantum computers, they can achieve up to exponential speedup \cite{Ardle2020, Bauer2020}. However, disciplines such as machine learning, finance, cryptography, optimization, image processing, and linear algebra can also benefit from quantum advantage \cite{Rebentrost2014, Schuld2017, Li2020, Orus2019, Woerner2019, Stamatopoulos2020, Martin2021, Ichikawa2023}. Many of these algorithms target the desired state for which multi-controlled (MC) gates are used. The most famous one is Grover's search algorithm \cite{Grover1996} that utilizes MC gates in diffusion operator. For that reason, efficient implementation of MC gates becomes essential for applying these algorithms to the genuine quantum device.

A systematic approach for decomposing multi-controlled gates is described in Ref.~\cite{Barenco1995}. This approach was subsequently optimized by alternative implementations, such as using $C-R_x$ rotations \cite{Saeedi2013, Silva2022}, partial removal of some gates from the circuit at the price of phase relativization \cite{Saeedi2013, Maslov2016}, or using ancilla qubits \cite{Balauca2022}. It is worth noting that the multi-controlled X (MCX) gate implementation can be straightforwardly generalized to implement an arbitrary multi-controlled single-qubit unitary gate \cite{Barenco1995,Nielsen2010}.

Our approach is inspired by the quantum Fourier transform (QFT) \cite{Nielsen2010} and its approximate (AQFT) form \cite{Barenco1996}. It was shown that many useful arithmetic operations may be approximated using the QFT \cite{Draper2000, Perez2017, Yuan2023}. We will extend this set by MCX gates. The paper is organized as follows: section \ref{sec2} provides a brief description of theoretical preliminaries for MCX implementation, section \ref{sec3} gives implementation details and methods for optimization using ancilla qubits, and section \ref{sec4} concludes the paper. The proof of equivalence of our implementation and the standard one is given in Appendix \ref{secA}.

\section{Theoretical preliminaries}\label{sec2}

When viewed in the computational $Z$-basis, an $n$-qubit multi-controlled-NOT gate performs a conditional $\pi$-rotation on the most significant (target) qubit around the $x$-axis based on the state of $n_c=n-1$ lower qubits. We should note that the choice of basis states is not unique since any two orthogonal vectors of a qubit system may be used as the computational basis states. Nonetheless, using the proper set of rotations, the computational basis can be converted to the desired. 

For simplicity, we will work with the qubits $\vert a_k\rangle$ in a pure state. Then $\vert a\rangle=\vert a_{n}\rangle \otimes\vert a_{n_c}\rangle=\vert a_{n}\rangle\otimes\vert a_{n-1}\rangle\otimes...\otimes\vert a_2\rangle\otimes\vert a_1\rangle\in Z_{2^n}$ is related to the binary representation of integer number $a$, where integer $a_{n_c}=a-a_{n}\cdot2^{n-1}$. In the classical case, an arithmetic operation performed on the integer $a$ stored in the $n$-bit register will update it to a value that is congruent modulo $2^n$. Increment of $a$ by one results in the value of the highest bit given by the Boolean expression $\overline x^{n_c}=a_{n}^{C^{n_c}X}=(a_1\cdot a_2 \cdots a_{n-1})\oplus a_{n}$, where the lower $n_c$ bits upgrade to a binary representation of $a_{n_c}+1\mod 2^{n_c}$. Applying a subsequent decrement by one only to the lower $n_c$ bits restores them to initial values, where the highest bit remains unchanged, thus storing the result of the controlled-NOT operation. From the standpoint of quantum computation viewed in the $\{\vert 0\rangle,\vert 1\rangle\}$-basis, this operation corresponds to the $n_c$-controlled $X$ operation, usually denoted by $C^{n_c}X$. Therefore, we will use this simple logic for an alternative MCX gate. 

The method employed for addition computes QFT on the first addend and then, based on the second addend, evolves it into QFT of the sum. Applying the inverse QFT recovers the sum to the computational basis. If we want to add a priory known classical value to quantum data, we only have to implement a series of corresponding rotations \cite{Draper2000}. To implement increment/decrement by one, we use a small set of rotations that can be executed simultaneously in a single time slice. The $QFT_n$ maps $n$-qubit state $\vert a\rangle$ to:
\begin{equation}
	QFT_n\vert a\rangle = \vert \widetilde a\rangle = \frac{1}{\sqrt{N}} \sum_{k=0}^{N-1} e^{i 2\pi a k/N}\vert k\rangle=\vert \phi_{n}(a)\rangle\otimes \vert \phi_{n-2}(a)\rangle\otimes\cdots\otimes\vert\phi_{1}(a)\rangle,\label{eqQFT}
\end{equation}
where $N=2^n$ and 
\begin{equation}
	\vert\phi_{k}(a)\rangle=\frac{1}{\sqrt{2}}(\vert 0\rangle+e^{i2\pi a/2^{k}}\vert 1\rangle).\label{eqPhik}
\end{equation}
The Fourier transform is carried through applying the Hadamard gate to each qubit $k$, followed by a sequence of two-qubit $C_j-R_{k,k-j+1}$ gates that conditionally apply rotations $R_{k-j+1}$ to the $k^{\rm th}$ qubit based on lower qubits ($j<k$) \cite{Draper2000}. A rotation of a single qubit by an angle $2\pi/2^m$, where $m\in Z$, is represented by the matrix:
\begin{equation}
	R_m=\begin{bmatrix} 1 & 0 \\ 0 & e^{i2\pi/2^m} \end{bmatrix}.\label{eqRm}
\end{equation}
An increment/decrement of $a$ by one in the transform range is achieved by applying an unconditional rotation $R_m$ to each qubit $m$. The $n$-qubit phase gate acting on $\vert\widetilde a\rangle$ is:
\begin{equation}
	P_{\pm1,n}\vert \widetilde a\rangle=R^{\pm 1}_n\otimes\cdots\otimes R^{\pm 1}_1 \vert \widetilde a\rangle=\vert \widetilde{a}\pm \widetilde{1} \:{\rm{mod}}\:{\widetilde{N}} \rangle.\label{eqPhase}
\end{equation}
The result $\vert \widetilde{a}\pm \widetilde{1}\:\:{\rm{mod}}\:{\widetilde{N}} \rangle$ is returned to the computational basis using the inverse of the $QFT_n$ ($IQFT_n$):
\begin{equation}
	IQFT\vert \widetilde{a}\pm \widetilde{1}\:{\rm{mod}}\:{\widetilde{N}} \rangle  = \frac{1}{\sqrt{N}} \sum_{k=0}^{N-1} e^{-i 2\pi (a\pm 1) k/N}\vert k\rangle= \vert a\pm 1\:{\rm{mod}}\: N\rangle.\label{eqIQFT}
\end{equation}
To perform the $n_c$-controlled $X$ operation, we increment the $n$-qubit data by one, obtaining the desired result stored in the highest qubit. Then, we restore the control qubits to their original value (``uncomputation") by carrying out a decrement by one on the data stored in the register composed of the lower $n_c$ qubits. The complete sequence is as follows:
\begin{align}
	\vert a\rangle &\xrightarrow[]{QFT_n} \vert \widetilde a\rangle \xrightarrow[]{P_{+1,n}} \vert \widetilde{a} + \widetilde{1} \:{\rm{mod}}\:{\widetilde{N}} \rangle \xrightarrow[]{IQFT_n} \vert a + 1\:{\rm{mod}}\: N\rangle \nonumber\\
	&=\vert \overline x^{n_c}\rangle\otimes \vert a_{n_c}+1\:\: {\rm{mod}}\:\frac{N}{2}\rangle\nonumber\\
	&\xrightarrow[]{I\otimes QFT_{n_c}} \vert \overline x^{n_c}\rangle\otimes\vert \widetilde a_{n_c}+\widetilde 1\: {\rm{mod}}\:\frac{\widetilde{N}}{2}\rangle \nonumber\\
	&\xrightarrow[]{I\otimes P_{-1,n_c}} \vert \overline x^{n_c}\rangle\otimes\vert \widetilde a_{n_c}\: {\rm{mod}}\:\frac{\widetilde{N}}{2}\rangle \nonumber\\
	&\xrightarrow[]{I\otimes IQFT_{n_c}} \vert \overline x^{n_c}\rangle\otimes\vert a_{n_c}\rangle. \label{eqAlg}
\end{align}

Although our implementation seems quite different from the standard MCX circuit, it is essential to note that they share the same unitary operator, as shown in the Appendix \ref{secA}.

\section{Implementation and discussion}\label{sec3}

\begin{figure}[h]
\centering
\includegraphics[width=0.9\textwidth]{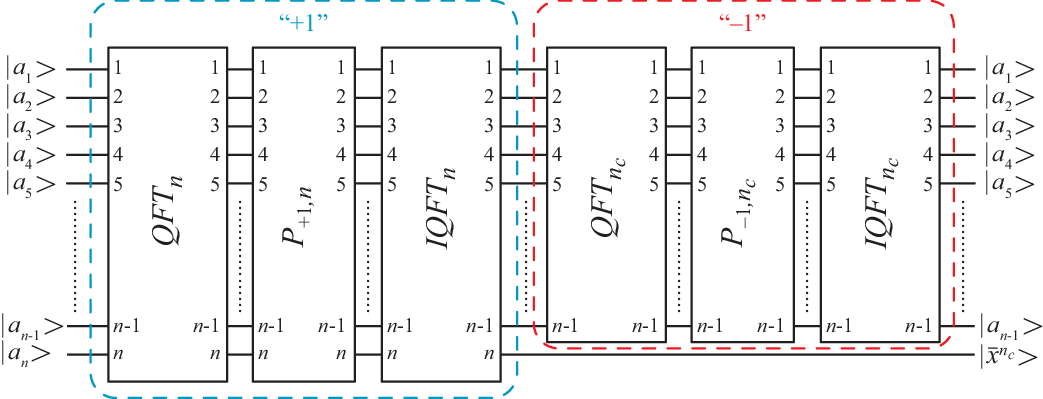}
\caption{A schematic diagram of $n$-qibit QFT-based MCX gate. QFT, P, and IQFT denote the quantum Fourier transform (Eq.~(\ref{eqQFT})), the phase gates (Eq.~(\ref{eqPhase})), and the inverse of QFT (Eq.~(\ref{eqIQFT})), respectively. The part of the quantum circuit labeled {\color{RoyalBlue}``+1''}/{\color{red}``$-$1''} for increment/decrement is framed by a {\color{RoyalBlue}blue}/{\color{red}red} dashed line.}\label{fig1}
\end{figure}

To implement circuits, Qiskit (ver. 0.42.1) \cite{Qiskit} is used. Initially, we analyze circuits that do not require ancilla qubits. Eq.~\ref{eqAlg} gives details on the gates needed to implement an MCX. A schematic diagram of the $n$-qubit MCX circuit is displayed in Fig.~\ref{fig1}. First, the QFT is appended to the $n$-qubit input. The ordering of control qubits in the $QFT_n$ circuit is irrelevant, but the most significant qubit should be the controlled (output) qubit. In succeeding subcircuits, this arrangement of qubits must be respected. The increment by one is achieved by applying a series of phase gates, as described by Eq. \ref{eqPhase}. Since these phase gates act on different qubits, they should be executed simultaneously, that is, at a single time slice. The IQFT reverts the incremented value to the computational basis, thereby setting the result of the MCX operation in the most significant qubit. The operator, that executes the increment by one, is given by the $IQFT_n\: P_{+1,n}\: QFT_n$ (it is labeled by {\color{RoyalBlue}''+1''} and framed by a {\color{RoyalBlue}blue} dashed line in Fig.~\ref{fig1}). Then, we apply the inverse operator to the control qubits $IQFT_{n_c}\: P_{-1,n_c}\: QFT_{n_c}$ (labeled by {\color{red}''-1''} and framed by a {\color{red}red} dashed line in Fig.~\ref{fig1}). It carries out the decrement by one, thus restoring their initial value. It is important to note that the proper ordering of qubits for this inverse operation should match the ordering of the lowest $n_c=n-1$ qubits.

\begin{figure}[h]
\centering
\includegraphics[width=0.88\textwidth]{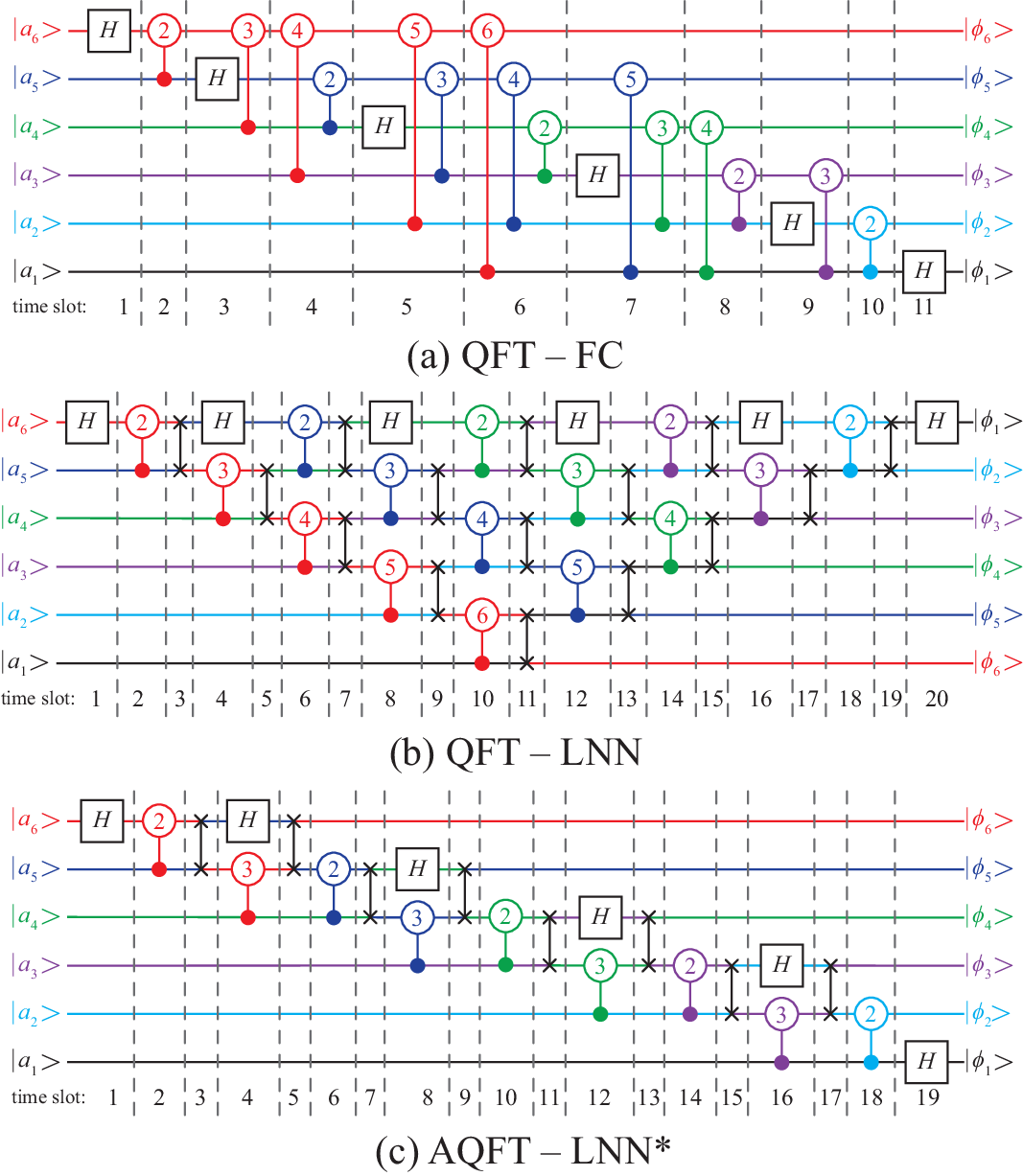}
\caption{A schematic diagram of the six-qubit (a) QFT in the fully connected (QFT-FC), (b) QFT in the linear-nearest-neighbor (QFT-LNN), and (c) optimized AQFT in the LNN (AQFT-LNN*) quantum computer architecture. Controlled $C_j-R_{j\prime,m}$ rotations are denoted with a line pointing from the control $j^{\rm th}$ qubit to the target $j\prime^{\rm th}$ qubit, where it ends with a circle. This circle indicates rotation gate $R_m$ with index $m$ inscribed in it. To make it easier to visualize, rotational gates applied to different qubits have different colors. Gates are divided into time slots by vertical gray dashed lines, where the index of the time slot is explicitly denoted at the bottom. The standard procedure to get approximate forms of circuits displayed in panels (a) and (b) is to remove controlled rotations denoted by index $m>3$.}\label{fig2}
\end{figure}

The number of gates needed for a standard QFT is $n(n+1)/2$ \cite{Nielsen2010, Draper2000}. After closely examining the schematic diagram of the QFT \cite{Draper2000}, it becomes clear that every wireline starts with the Hadamard gate, followed by a series of conditional rotations controlled by the lower-order qubits that have not yet undergone the Hadamard gate. This setup imposes specific restrictions for parallel execution. The diagram in Fig.~\ref{fig2}(a) shows the six-qubit QFT with clustered gates that can be executed simultaneously in a quantum computer architecture that supports interaction between arbitrary pairs of qubits. We will be using the term ``fully-connected" (FC) when referring to this architecture. Only gates acting on different qubits can be executed at a single time slice. To perform the QFT in the FC architecture (QFT-FC), each wireline $\vert a_k\rangle$, except for the one carrying the most significant qubit, must be appended with the $C_k-R_{k+1,2}$ gate. Thereby, the $k^{\rm th}$ qubit will control the $R_2$ rotation applied to the $(k+1)^{\rm th}$ qubit. Once this is done, the Hadamard gate can act on the $k^{\rm th}$ qubit. Notice that these two gates cannot be executed simultaneously as both act on the same $k^{\rm th}$ qubit. In the single odd time slice, we group the Hadamard gate acting on the $k^{\rm th}$ qubit with $C_{k-p}-R_{k+p,2p+1}$ gates, where $0<p\leq\min\{k,n-k\}$. The subsequent even time slice concentrates $C_{k-q}-R_{k+q-1,2q}$ gates, where $0<q\leq\min\{k-1,n-k+1\}$. It is straightforward to conclude that at most $\left\lceil n/2\right\rceil$ gates will be executed simultaneously and that the QFT can be performed in $f_{t.slots}^{QFT}(n)=2n-1$ time slots, provided that the quantum hardware implements an arbitrary controlled rotation as a basic operation. This number of time slots is the lower bound for the QFT's time complexity.
From Fig.~\ref{fig2}(a), one may find that QFT uses $f_H^{QFT}(n)=n$ Hadamard gates and $f_{C-R}^{QFT}(n)=n(n-1)/2$ controlled rotations. The total number of non-elementary gates is $f_{gates}^{QFT}(n)=n(n+1)/2$. To implement MCX in FC architecture, we use $n$-qubit and $(n-1)$-qubit  $QFT-P-IQFT$ block of gates. The number of $R_m$ gates in P is $f_{gates}^P(n)=n$ and all can be executed simultaneously. The total number of non-elementary gates used is $f_{gates}^{MCX-FC}(n)=2*f_{gates}^{QFT}(n)+f_{gates}^P(n)+2*f_{gates}^{QFT}(n-1)+f_{gates}^P(n-1)=2n^2+2n-1$. The number of time slots for $n$-qubit MCX execution is $f_{t.slots}^{MCX-FC}(n)=2*f_{t.slots}^{QFT}(n)+2*f_{t.slots}^{QFT}(n-1)+2=(8n-6)$.


To optimize circuit complexity one may want to consider using an approximate QFT (AQFT). From Eq.~\ref{eqRm}, we infer that the rotation angle gets very small as $m$ gets large. Thus, the rotation matrix $R_m$ approaches the identity matrix. Additionally, it is crucial to consider whether these rotation gates can be executed with a certain level of tolerance in practical applications. Without such precision, the effectiveness of these gates can be in question. Studies have shown that a truncated QFT may provide greater accuracy than a full QFT in the presence of decoherence \cite{Barenco1996}. The optimal value for $m$ needed is estimated to $[\log_2 n]$. This will reduce the number of essential operations required to $f_{gates}^{AQFT}=(2n-[\log_2 n])([\log_2n]-1)/2\approx n[\log_2 n]$ \cite{Barenco1996, Draper2000}, while at most $\left\lceil [\log_2n]/2\right\rceil$ gates will be executed simultaneously. Therefore, using AQFT will not reduce the minimum circuit depth but only the total number of gates needed and the maximum number of gates executed simultaneously. Thus, $(8n-6)$ is the lower bound for the time complexity estimation. The total number of non-elementary gates used for AQFT-based $n$-qubit MCX is $f_{gates}^{AMCX-FC}(n)=2*f_{gates}^{AQFT}(n)+f_{gates}^P(n)+2*f_{gates}^{AQFT}(n-1)+f_{gates}^P(n-1)$ which is the lower bound for the space complexity calculation.

Still, in many architectures, each qubit has a finite number of neighbors bounded by a number $l$ \cite{Fowler2004}. To apply conditional rotation between two qubits, they must be neighbors. If $[\log_2n]>l$, we need to add SWAP gates, which increases the circuit depth. In recent studies, the lower bound of time slices needed for the QFT execution is estimated to be $[(2+2/l)n+O(1)]$ \cite{Maslov2007}. The most restrictive case is for the linear-nearest-neighbor (LNN) computer architecture, where two-qubit gates are allowed only between qubits whose subscript values differ by one ($l=2$). To make qubits nearest neighbors, we must swap some pairs of qubits before controlled rotation can be applied. A schematic diagram of the six-qubit QFT in the LNN architecture (QFT-LNN) is displayed in Fig.~\ref{fig2}(b). Following the one wireline, only the first rotation $C_{n-1}-R_{n,2}$ gate can be applied without swapping qubits since acting on the nearest neighbors. The following controlled rotation should be executed on the next-nearest-neighbors, so we swap $n^{\rm th}$ and $(n-1)^{\rm th}$ qubit. Therefore, formerly $(n-1)^{\rm th}$ and $(n-2)^{\rm th}$ qubits are no longer neighbors, so we have to perform another swap before the second set of rotations, and so on. Notice that each set of elementary gates executed at a single time slice in Fig.~\ref{fig2}(a) has a corresponding pair of time slices in Fig.~\ref{fig2}(b) that are used to swap qubits and perform the equivalent set of operations, respectively. Consequently, we will have additional $f_{t.slots}^{SWAP(QFT)}(n)=(2n-3)$ time slots for swapping qubits with $f_{gates}^{SWAP(QFT)}(n)=(n-1)(n-2)/2$ SWAP gates. Depending on the basis used for calculation, each SWAP gate is implemented using a few basis gates. Notice that qubit ordering is reversed. When implementing MCX, instead of using additional $\lfloor n/2\rfloor$ SWAP gates at QFT(IQFT) output, the subcircuits succeeding a QFT or IQFT should be appended with inputs swapped. Alternatively, QFTs and IQFTs inputs should have reverse ordering. To implement $n$-qubit MCX in LNN architecture, the number of time slots required is $f_{t.slots}^{MCX-LNN}(n)=f_{t.slots}^{MCX-FC}(n)+2*f_{t.slots}^{SWAP(QFT)}(n)+2*f_{t.slots}^{SWAP(QFT)}(n-1)=16n-22$ which is the upper bound for the time complexity estimation. The number of non-elementary gates is $f_{gates}^{MCX-LNN}(n)=f_{gates}^{MCX-FC}(n)+2*f_{gates}^{SWAP(QFT)}(n)+2*f_{gates}^{SWAP(QFT)}(n-1) = 4n^2-10n+7$ which will be the upper bound to estimate the space complexity.

Again, the question is whether the circuit can benefit from approximation. The straightforward procedure to obtain an AQFT in the LNN (AQFT-LNN) is to remove controlled rotations with an index $m>[\log_2n]$, which will reduce the number of controlled phases as for the AQFT-FC. The circuit will have the depth and the number of SWAP gates used as the QFT-LNN. Attempting to reduce the number of SWAP operations will increase the depth of the circuit in return. Therefore, we consider the former method optimal. The exception is $n<6$ when we perform only $C-R_2$ gates on the adjacent qubits. Thus, we do not need SWAP gates, so the AQFT-FC and AQFT-LNN will match. The second exception is $6\le n<12$ when we need $C-R_2$ and $C-R_3$ gates. To perform consecutive $C-R_2$ and $C-R_3$ rotations, we have to swap one of the next-nearest-neighbors qubits to do $C-R_3$ rotation and then swap back to restore qubit ordering for the following $C-R_2$ rotation. A schematic diagram of the optimized six-qubit AQFT in the LNN architecture (AQFT-LNN*) is displayed in Fig.~\ref{fig2}(c). This implementation is less complex than the AQFT-LNN obtained from the circuit in Fig.~\ref{fig2}(b) by removing controlled rotations denoted by $m>3$. Moreover, qubits ordering is not reversed.

Up to this point, we assumed that a quantum computer inherently implements the single-qubit $R_m$ and the two-qubit $C-R_m$ rotation, thus making them part of the universal gate set. In a genuine quantum device, this basis set (that is, the native gate set - NGS) is formed from the single- and two-qubit gates that have high fidelity and are usually the best choice for standard quantum circuit implementations. For example, superconducting hardware uses the single-qubit $R_z$ gate and the two-qubit $C-X$ gate rather than $R_m$ and $C-R_m$. 

\begin{figure}[h]
\centering
\includegraphics[width=0.9\textwidth]{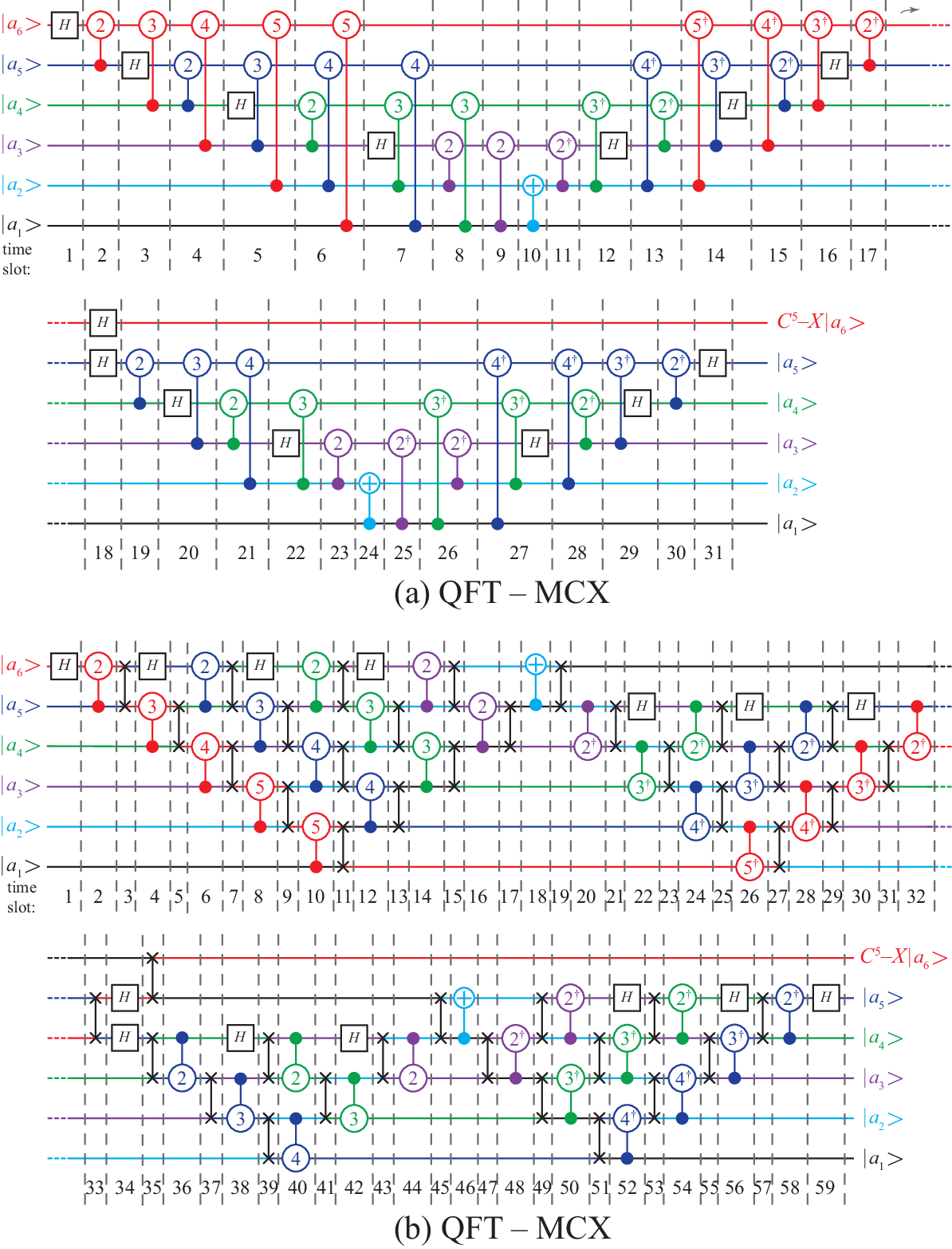}
\caption{Optimized QFT-based MCX in (a) FC, and (b) LNN architecture.}\label{fig3}
\end{figure}

A realistic assessment of the time and space complexities can be found from the circuit decomposed in the NGS. Detailed optimization and decomposition are elaborated in Appendix \ref{secB}. An optimized MCX circuits in the FC and LNN architectures are shown in Fig.~\ref{fig3}(a) and (b), respectively. Due to the merging controlled-phase and increment/decrement phase gates, the circuit reduces to $(8n-17)$ time slots in the FC architecture, as explained in Appendix \ref{secB}. At most $(8n-20)$ additional SWAP-gate slots are required for implementation in a qubits network with linear connectivity, so $(16n-37)$ time slots are needed in the LNN architecture. Counting in the NGS, the depth of MCX is $(32n-80)$ in the FC and $(56n-146)$ in the LNN (for $n>3$). For $n=3$ we use at most two SWAP gates in the LNN. If the target wireline can be placed between control wirelines in the Toffoli gate, we can avoid using SWAP gates. Therefore, the time complexity of the Toffoli gate in the FC and LNN will be the same.

FC-MCX uses $(4n-10)$ Hadamard gates, $(2n^2-6n+3)$ controlled phase gates, and two $C-$NOTs. The number of elementary gates in $H$ and $C-R_m$ is 3 and 5, respectively. Two $R_z$ gates annihilate between $C_j-R_m$ and $C_{j-1}-R_m^\dagger$ acting on the same target qubit, as discussed in Appendix \ref{secB}. Therefore, we will use $2(n-2)$  elementary gates less in {\color{RoyalBlue}``$+1$''}  and $2(n-3)$ in {\color{red}``$-1$''}. The total number of elementary gates is $(10n^2-22n-5)$. If we simplify the circuit using approximate QFT, the best solution is to keep $C-R_m$ gates with $m$ value up to $[\log_2n]$ after merging controlled phase gates in QFT and IQFT. Thereby, there will be a few more $C-R_m$ gates than if we approximate QFT, IQFT, and P separately. This way, the approximation will be better. Thus, we use approximate {\color{RoyalBlue}``$+1$''} and {\color{red}``$-1$''} circuits rather than approximate QFT(IQFT).  The number of $C-R_m$ gates will reduce to $2([\log_2n]-1)(2(n-1)-[\log_2n])$ for $n>3$, and the Toffoli gate will use only three $C-R_2$ gates. Approximate MCX in the FC architecture will use $10([\log_2n]-1)(2(n-1)-[\log_2n])+10n-23$ elementary gates which is the space complexity lower bound.

LNN-MCX uses at most $(2n^2-6n+6)$ swap gates. In adjacent SWAP and $C-$NOT, two $C-$NOTs annihilate. Therefore, the number of $C-$NOTs used for swapping is $6n^2-18n+14$. Assuming that there must be a defined arrangement (ascending or descending) of qubits, the number of elementary gates used in the LNN architecture is $(16n^2-40n+9)$, which is the space complexity upper bound. If the approximate form is used, the number of elementary gates reduces to $10([\log_2n]-1)(2(n-1)-[\log_2n])+6n^2-8n-9$.

\begin{figure}[h]
\centering
\includegraphics[width=0.95\textwidth]{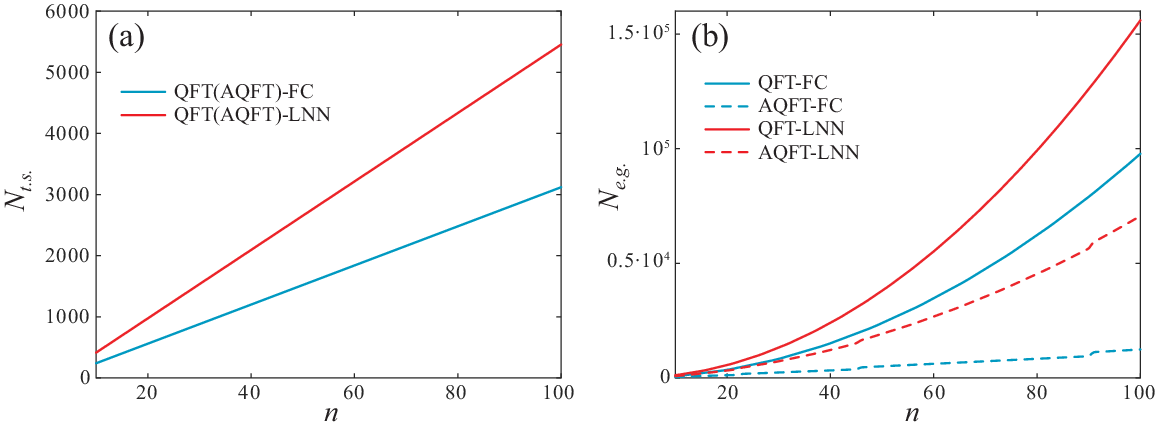}
\caption{(a) The number of time slices ($N_{t.s.}$), and (b) the number of elementary gates ($N_{e.g.}$) as a function of the number of qubits in the MCX circuit implemented in the fully connected (FC - {\color{RoyalBlue}blue} line) and the linear-nearest-neighbor (LNN - {\color{red}red} line) architecture. Solid (dashed) lines are used for the QFT(AQFT)-based MCX circuit.}\label{fig4}
\end{figure}

To compare the time and space complexities of different implementations, we plot the circuit depth and the number of gates needed as a function of the number of qubits in Figs.~\ref{fig4}(a) and (b), respectively. For all considered implementations, the number of the time slices depends linearly on the number of qubits in the MCX, as displayed in Fig.~\ref{fig4}(a). QFT-based and AQFT-based implementations are executed in an equal number of time slices. However, the time slice count depends on the number of SWAP gates needed for implementation in a particular computer architecture. The LNN uses the most SWAP gates, while FC architecture doesn't require any. The number of time slices needed to execute MCX in the FC and LNN set the lower and the upper bound for the time complexity, respectively.

Also, we compare the number of elementary gates ($N_{e.g.}$) required for different implementations. For the QFT-based $n$-qubit MCX, there is a square increase in the number of elementary gates with $n$ for both FC and LNN implementation, as displayed by solid lines in Fig.~\ref{fig4}(b). The QFT-LNN has a higher positive slope since there is also an additional square increase in the number of SWAP gates with $n$. This implementation uses the highest number of elementary gates and is considered the upper bound in the gates count. The number of rotation gates in AQFT-based implementations is a function of $[\log_2n]$, which is a stair-step function. It is reflected in a stair-wise dependence of the number of elementary gates on $n$. This is easy to comprehend because there is a steep change in the number of controlled phase gates used in an AQFT with $[\log_2n]$. Since there is no need for SWAP gates in the AQFT-FC, the number of elementary gates is the smallest, thus setting the lower bound. The AQFT-LNN implementation additionally uses $O(n^2)$ SWAP gates, which result in a larger count of elementary gates to the number of rotation gates required. Therefore, it is hard to notice that the increase in $N_{e.g.}$ is stair-wise, which can be inferred from the dashed {\color{red}red} line in Fig.~\ref{fig4}(b). This differs from the AQFT-FC (dashed {\color{RoyalBlue}blue} line), where $N_{e.g.}$ is a linearly increasing piecewise function with $n$.

\begin{figure}[h]
\centering
\includegraphics[width=0.95\textwidth]{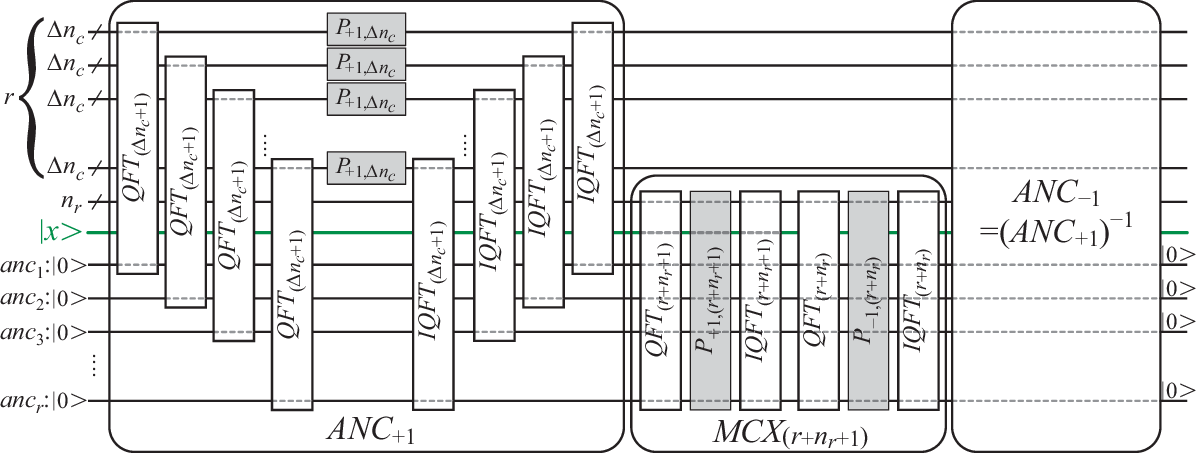}
\caption{A schematic diagram of QFT-based MCX circuit using $r$ ancilla qubits. The left part of the circuit, labeled by $ANC_{+1}$, performs increment by one on sets of $\Delta n_c$ controlled qubits and one ancilla qubit. In this step, the values of r registers each containing $r\Delta n_c$ control qubits are incremented, while ancillae are used as the ``carry out" qubits. All $QFT_{(\Delta n_c+1)}$ are run in parallel, thus limiting the number of time slices to the value needed for executing one $QFT$. The same parallelization applies to IQFTs, while all phase gates can be executed in a single time slice. The central part of the circuit, denoted by $MCX_{r+n_r+1}$, performs $X(NOT)$ operation to the target {\color{ForestGreen}$\vert x\rangle$} qubit controlled by the group of $r$ ancilla qubits and $n_r$ remnant control qubits. The right part of the circuit, $ANC_{-1}$, is the inverse of $ANC_{+1}$. It executes decrement by one to groups of joined $\Delta n_c$ control and ancilla qubits, thus restoring ancillae to their initial state $\vert 0\rangle$. In each subcircuit schematic block ($QFT$, $IQFT$, or $P$), dashed gray lines indicate which qubits are used.}\label{fig5}
\end{figure}

Low quantum circuit complexity can lead to more efficient quantum computation that is less prone to errors. One way to reduce the depth of a circuit is by utilizing ancilla qubits. Suppose we have a certain number ($r$) of ancilla qubits that are in the initial state $\vert 0 \rangle$ (see Fig.~\ref{fig5}). We divide the control qubits into $r$ equal groups, each containing $\Delta n_c$ qubits. In the general case, we may have $n_r = n_c-r\cdot \Delta n_c$ extra control qubits, which we'll address later. For each group, we use QFT-based ``increment by one circuit'' on $\Delta n_c$ qubits and one ancilla as the ``carry out'' qubit. The operation will switch ancilla qubit to $\vert 1 \rangle$ if all $\Delta n_c$ qubits are also in the state $\vert 1 \rangle$. These increment operations are executed in parallel, applying all $QFT_{\Delta n_c+1} (IQFT_{\Delta n_c+1})$ simultaneously and executing $P_{+1,\Delta n_c+1}$ at a single time slice (note that we do not need to implement $P_{\pm1}$ separately due to the optimization described in Appendix \ref{secB}). Therefore, dividing control qubits into equal-sized groups is optimal, which was our initial assumption. These operations, which perform switching the state of ancilla qubit to $\vert 1 \rangle$ if the state of $\Delta n_c$ control qubits is $\vert 1 \rangle$, are grouped in a single block labeled $ANC_{+1}$, as displayed in Fig.~\ref{fig5}.

Afterward, we execute the operation controlled by the ancilla qubits and the remaining control qubits to the most significant qubit $\vert x\rangle$ (see Fig.~\ref{fig5}). Note that this MCX operation is implemented in a way described in the explanation regarding Fig.~\ref{fig1}. The group of subcircuits used for MCX is denoted by $MCX_{(r+n_r+1)}$ in the middle section of Fig.~\ref{fig5}. Finally, we apply the inverse of the $ANC_{+1}$ operation (denoted by $ANC_{-1}$), executing ``decrement by one" to $r$ batches of joined $\Delta n_c$ controls with each ancilla. Thereby, ancillae are restored to their initial state $\vert 0\rangle$. For successful ``uncomputation" of ancilla qubits, the ordering of qubits in $ANC_{+1}$ and $ANC_{-1}$ must match.

When using auxiliary qubits, it is crucial to find an optimal implementation. It is easy to comprehend that multiple ancilla qubits allow independent parts of the algorithm to be executed simultaneously. In our case, parallelization is possible in $ANC_{\pm1}$. Therefore, we divide the algorithm into two parts. One part is doing $ANC_{+1}$ and $ANC_{-1}$ with groups of control qubits that we can split among ancilla qubits and process them in parallel, while the other computes MCX operation ($MCX_{r+n_r+1}$) on the remaining control qubits, the ancilla qubits, and one ``MCX output'' qubit (denoted by {\color{ForestGreen}$\vert x\rangle$} in Fig.~\ref{fig5}). Each part has to execute pairs of $QFT$, $IQFT$, and $P$ gates. The complexity of both parts is constrained by the QFT implementation. $MCX_{r+n_r+1}$ has a lower complexity than a single cluster in $ANC_{\pm1}$ of the same size since it doesn't have to ``uncompute'' the most significant qubit. 

First, we explore how the dividing of control qubits into clusters influences the circuit complexity. Increasing the number of control qubits in the cluster, the depth of the circuit will decrease. This is easy to follow since an increase of $\Delta n_c$ by one means the algorithm parallelized the processing of additional $r$ control qubits in $ANC_{\pm1}$. The depth of QFTs(IQFTs) in $ANC_{\pm 1}$ will enlarge due to only one additional qubit in each cluster. At the same time, there will be $r$ qubits less in MCX. Consequently, the depth of the circuit will linearly decrease with $\Delta n_c$. The minimum depth of the circuit is when the sum of $ANC_{\pm1}$ and $MCX_{(r+n_r+1)}$ depths is minimal, which is for:
\begin{equation}
	\Delta n_{c,{\rm{max}}}=\left[\frac{n_c}{r}\right].\label{eqDncmax}
\end{equation}
This is considered the maximum number of control qubits in a cluster.

However, the number of elementary gates depends nonlinearly on $\Delta n_c$. We will find the optimal cluster size to reach the minimum number of elementary gates building the circuit. When using $r$ ancilla qubits, we form $r$ groups containing $\Delta n_c$ control qubits plus one ancilla qubit to which we apply increment or decrement operations in parallel. Thereby, we are dealing with $r\Delta n_c + r=n_c-n_r+r$ qubits in $ANC_{\pm1}$. The remaining $n_r$ control qubits, $r$ ancilla qubits, and one the most significant qubits are used for executing $MCX_{(r+n_r+1)}$. It is straightforward to show that the minimum number of gates used is when MCX and a single cluster in $ANC_{\pm1}$ have approximately equal depths and numbers of elementary gates. To find the appropriate clusters size, which is $(\Delta n_c+1)$, we have to divide $((n_c - n_r +r)+(n_r + r+1))$ into $(r+1)$ approximately equal-sized groups. For the number of ancilla qubits constrained to $r$, provided we have to use them all, the optimal value for $\Delta n_c$ is:
\begin{equation}
	\Delta n_{c,{\rm{opt}}}=\left[\frac{n_c + r}{r+1}\right].\label{eqDnc}
\end{equation}

An increase in the number of ancilla qubits permits higher parallelization that will reduce execution time. We will find the minimum number of ancilla qubits that will provide an optimal circuit. If the number of auxiliary qubits is not constrained, we can split all control qubits to approximately equal size clusters when $n_c\approx r^2$, thus $r_{\rm{opt}}=[\sqrt{n_c}]$. In this case, $\Delta n_{c,{\rm{opt}}}$ and $\Delta n_{c,{\rm{max}}}$ are equal or differ by at most one that follows from Eqs.~(\ref{eqDncmax}) and (\ref{eqDnc}). The ideal scenario is when the number of control qubits has a whole number as its square root and if we assume that the number of ancilla qubits used is equal to the square root of the number of control qubits. As a consequence, $\Delta n_{c,{\rm{opt}}}(\sqrt{n_c})=\Delta n_{c,{\rm{max}}}(\sqrt{n_c})$.

For the number of auxiliary qubits above the optimal, control qubits are all divided into clusters where most count $\Delta n_{c,{\rm max}}^{r>r_{\rm opt}}= \lceil n_c/r\rceil$ control qubits, while others have one less. One may note that clusters in $ANC_{\pm1}$ are equal in size only when $n_c$ is divisible by $r$. The depth of $ANC_{\pm1}$ is determined by the size of a larger cluster $\Delta n_{c,{\rm max}}^{r>r_{\rm opt}}$ that is decreasing stair-step function. On the other hand, the depth of the MCX increases linearly with $r$ since we need to implement an $MCX_{(r+1)}$ circuit. Therefore, the time complexity will increase linearly except if the increase of $r$ by one results in the decrease of cluster size in $ANC_{\pm1}$ by one, when the depth does not change. 

The number of elementary gates to implement the QFT(IQFT) in MCX is proportional to $r^2$, so the number of gates needed for MCX is $O(r^2)$. The QFT(IQFT) in $ANC_{\pm1}$ uses approximately $O((\Delta n_{c,{\rm max}}^{r>r_{\rm opt}})^2)$ gates. Considering $r$ clusters in $ANC_{\pm1}$, the number of elementary gates decreases with $r$ as $O(n_c^2/r)$. Based on a simple analysis, we conclude that there is a relatively small variation in the number of elementary gates with $r$ when using a large number of ancillas.

\begin{figure}[h]
\centering
\includegraphics[width=1\textwidth]{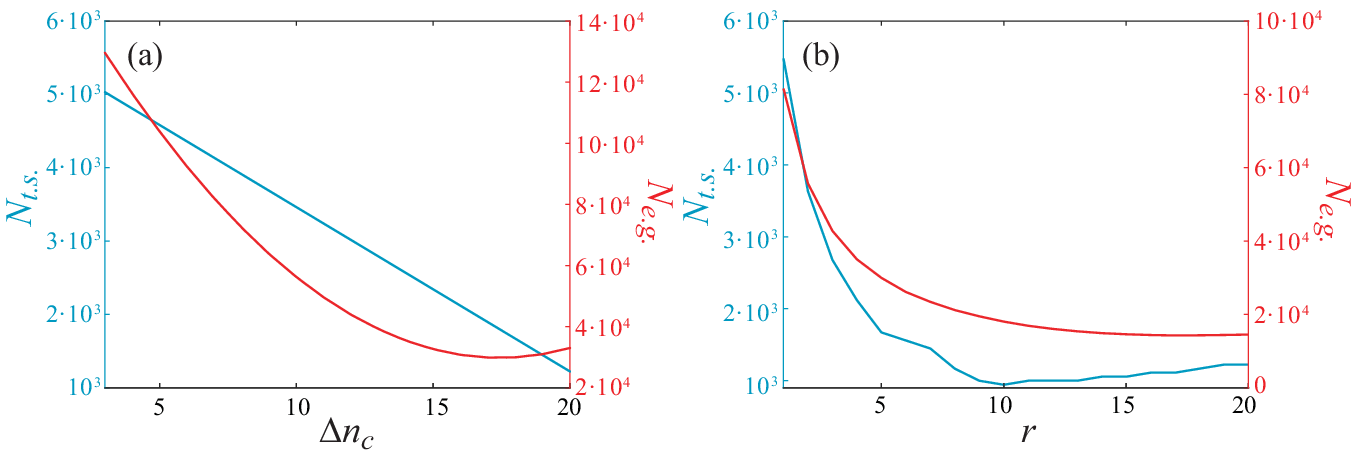}
\caption{The dependence of the number of time slices $N_{t.s.}$ (left axis, {\color{RoyalBlue}blue} line) and the number of elementary gates used $N_{e.g.}$ (right axis, {\color{red}red} line) on (a) the number of control qubits in a cluster $\Delta n_c$ using five ancillas and (b) the number of auxiliary qubits $r$ using the optimal cluster size given by Eq.~(\ref{eqDnc}), in the 101-qubit QFT-based MCX.}\label{fig6}
\end{figure}

In various architectures used for quantum computing, ranging from the most efficient FC to the most restricted LNN, there is a similar change in circuits complexity with the number of ancilla qubits used. We will analyze only the most complex QFT-LNN-based implementation of the 101-qubit MCX. In Fig.~\ref{fig6}(a) we show the dependence of the number of time slices ($N_{t.s.}$) and elementary gates used ($N_{e.g.}$) on the number of control qubits in a cluster ($\Delta n_c$) in the MCX that uses $r=5$ auxiliary qubits. Considering the number of control and ancilla qubits, a cluster can contain at most $\Delta n_{c,{\rm max}}=20$ control qubits. One may notice a linear decrease in the number of time slices with $\Delta n_c$, where the minimum time complexity is reached for the maximum cluster size (see {\color{RoyalBlue}blue} line in Fig.~\ref{fig6}(a)). However, the dependence of $N_{e.g.}$ on the $\Delta n_c$ is nonlinear (see {\color{red}red} line in Fig.~\ref{fig6}(a)), as explained afore. The minimum number of gates used is for $\Delta n_c=17$, which also follows from Eq.~(\ref{eqDnc}). 

The dependence of time slices and elementary gates needed on the number of ancilla qubits ranging from 1 to 20 is shown in Fig.~\ref{fig6}(b). Both $N_{t.s.}$ and $N_{e.g.}$ start decreasing as $r$ increases, as indicated by the {\color{RoyalBlue}blue} and {\color{red}red} line in Fig.~\ref{fig6}(b), respectively. For $r=r_{opt}=\sqrt{n_c}=10$, there is a minimum of time slices while the slope of $N_{e.g.}$ becomes very small.  For $r>r_{\rm opt}$, the depth linearly increases with $r$ except for values when cluster size in $ANC_{\pm1}$ decreases by one. This is manifested in narrow horizontal lines in $N_{t.s.}$, which is evident in Fig.~\ref{fig6}(b). On the other hand, the number of elementary gates required slowly decreases as $r$ increases above $r_{\rm opt}$. To conclude, an increase in $r$ above the optimal value diminishingly influences a decrease in the number of elementary gates required at the price of increasing time slices needed for execution and using an unnecessarily large number of auxiliary qubits. Therefore, we consider $r_{\rm opt}$ the best number of auxiliary qubits to use.

Since the QFT is an essential subcircuit used in our approach, one may wonder if there is another way to improve its efficiency other than AQFT. The state-of-the-art optimization \cite{Park2023} shows that the number of CNOT gates used in the LNN QFT can be reduced to about $40\%$ compared to previously known LNN QFT. Nevertheless, this approach increases circuit depth which cancels the benefit of a decrease in the number of elementary gates. If we can use an arbitrarily large number of ancilla qubits, we might consider parallelizing QFT, as discussed in Ref.~\cite{Cleve2000}. This approach imposes the upper bound of $O(\log_2n)$ for the time complexity and $O(n(\log_2n)^2\log_2\log_2n)$ on the circuit size in $n$-qubit QFT. Although this method significantly reduces circuit depths, it considerably increases the number of qubits used making it less scalable for implementation in current noisy intermediate-scale quantum (NISQ) devices. Therefore, we consider AQFT to be the optimal choice.

The possibility for simplifying three-qubit controlled gates, if arbitrary phase shifts of qubit states are permitted, was first elaborated in Ref.~\cite{Barenco1995}. This approach is acceptable if the gate is part of an operation that straightforwardly maps some classical mathematical expression to reversible computation or if gates can be arranged to cancel out this extra phase. In general nonclassical unitary operation, this extra phase is ``dangerous''. Saeedi and Pedram first recognized the advantages of using a relative phase quantum circuit for implementing $n$-qubit Toffoli gates \cite{Saeedi2013}. This approach is further elaborated in Ref.~\cite{Maslov2016}. Several recent state-of-the-art implementations are based on optimizing this approach \cite{Silva2022, Balauca2022, Jun2023}. All these implementations result in depths of MC gates that are linear on the number of qubits and use a relatively small number of elementary gates. The QFT-based implementation also has a linear depth and more importantly, does not exhibit phase relativization.

One of the most efficient and systematic procedures to decompose multi-controlled unitary gates (the linear-depth decomposition - LDD) is presented in Ref.~\cite{Silva2022}. The authors calculated the lower bound for the number of time slots for execution to be $(8n-20)$ in the FC architecture. It is one of the best implementations concerning the prediction of the MCX complexity. This is comparable to the number of time slots $(8n-17)$ in our approach. To implement $n$-qubit MCX the authors used $C-X^{1/2^{m-1}}=C-R_x(\pi/2^{m-1})$ gates as a part of the universal gate set. However, the decompositions of $C-R_x(\pi/2^{m-1})$ in the NGS use 11 (for $m=2$) and 12 (for $m>2$) elementary gates (two $C-X$, four $\sqrt{X}$ and the rest are $R_z$), respectively. Both gates need 11 time sequences to execute. This is approximately twice what we have for $C-R_m$. We infer that our approach can outperform this state-of-the-art one.

The authors demonstrated the proof-of-principle using IBM's quantum cloud platform. For LDD-MCX implementation ibm\_hanoi, a 27-qubit Falcon r5.11 processor, was used. One should note that our implementation also uses the Falcon's family native gate set. They demonstrated that the quantum circuit depth of their MCX implementation increases much more slowly with the number of qubits than in other known methods. Up to the 6 qubits, LDD-MCX yields relatively deep circuits. For a number of qubits above 6, their implementation almost exponentially increases the advantage compared to the standard one \cite{Silva2022}.

\begin{figure}[h]
\centering
\includegraphics[width=0.55\textwidth]{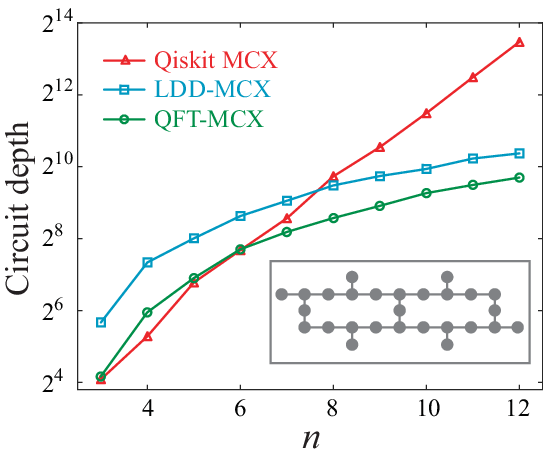}
\caption{The dependence of circuit depths on the number of qubits in {\color{red} default Qiskit} ({\color{red}$\bigtriangleup$}), {\color{RoyalBlue}LDD-MCX} from Ref.~\cite{Silva2022} ({\color{RoyalBlue}$\Box$}), and  {\color{ForestGreen}QFT-MCX} ({\color{ForestGreen}$\bigcirc$}). A schematic view of qubits connectivity in the ibm\_hanoi quantum processor is shown in the insert.}\label{fig7}
\end{figure}

We transpiled our circuit using the local simulator and compared circuit depths with state-of-the-art implementation publicly available on the GitHub page \cite{MCXcode2022}. This comparison is displayed in Fig.~\ref{fig7}. We used optimization level 3 in Qiskit's transpile function. In Ref.~\cite{Silva2022} optimization level 2 is used. A higher optimization level does not significantly affect the LDD-based and QFT-based implementations but results in a more optimal default Qiskit MCX circuit. As suggested above, our approach is approximately twice as efficient as the state-of-the-art one, which can be inferred by comparing the  {\color{ForestGreen}green} ({\color{ForestGreen}$\bigcirc$}-markers) and {\color{RoyalBlue}blue} ({\color{RoyalBlue}$\Box$}-markers) lines in Fig.~\ref{fig7}. Up to the 6-qubit MCX, the default Qiskit implementation is comparable to the QFT-based, but outperforms the LDD circuits. However, the default Qiskit implementation shows an exponential ($\sim 2^n$) increase in the circuit depth with the number of qubits ({\color{red}red} line with {\color{red}$\bigtriangleup$} markers in Fig.~\ref{fig7}). For the optimization level used, when the number of qubits $n>7$ in LDD-MCX or $n>6$ in QFT-MCX, both become significantly less complex than the standard implementation. Since the LDD-MCX and QFT-MCX are linear depth circuits, they both gain an exponential advantage over the default MCX above this critical number of qubits. A more detailed analysis can show that the complexity of the QFT-based MCX gates is close to the upper bound. This is due to relatively sparse qubit connectivity in superconducting hardware. The schematic view of qubit connectivity in ibm\_hanoi device is displayed in the insert of Fig.~\ref{fig7}.

Based on the LDD-MCX and QFT-MCX decompositions in the NGS, we concluded that the LDD-MCX uses twice the number of elementary gates as the QFT-MCX. These elementary gates have a high, but finite fidelity. Increasing the number of elementary gates used in a quantum computation increases an error in the computation process. Therefore, our circuit has a higher fidelity.

We should note that the LDD-MCX implementation in Ref.~\cite{Silva2022} does not use auxiliary qubits. This circuit can be simplified using ancilla qubits applying the method explained for the QFT-MCX. Since clusters in both implementations will be the same size (counting in a basis set with non-elementary gates), the advantage of our implementation will be preserved even when comparing the LDD-based and QFT-based implementations that use auxiliary qubits.

\section{Conclusions}\label{sec4}

In this paper, we proposed a new multi-controlled X (MCX) gate implementation based on the quantum Fourier transform (QFT). Circuit implementation that does not use auxiliary qubits is explained in detail. The most connected/restricted quantum computer architecture was analyzed to obtain the lower/upper bound for the time and space complexities. There is a linear increase in the number of time slices and a quadratic in the number of required elementary gates with the number of qubits in MCX. The number of gates significantly reduces when using the approximate implementation of the QFT, while there is no improvement in the time complexity of the circuit. 

When using ancilla qubits, a circuit is divided into two functional parts. One performs in parallel increment/decrement by one on clusters composed of control qubits and one ancilla, while the other executes MCX operation on remnant control qubits, ancillas, and one output qubit. For a fixed number of auxiliary qubits used, the time complexity linearly decreases with the control qubits count in equal-size clusters. Constraining the number of required elementary gates to the minimum, the expression for the optimal cluster size was found. If the number of auxiliary qubits is not constrained, the optimal number that should be used is equal to the rounded square root value of the number of control qubits. Above this optimal number of ancillae, the time complexity of circuits increases whilst the number of gates insignificantly reduces. 

Similar optimizations concerning the clustering of control qubits and selecting the proper number of ancilla qubits can be employed in other MCX implementations. Moreover, it was inferred that many exotic quantum gates can be implemented using quantum arithmetics. In cases where it is necessary to implement more complex arithmetic operations for a gate, the QFT-based approach can be of great advantage. 

We compared our implementation with the state-of-the-art one. QFT-based MCX has a smaller complexity and higher fidelity than the most efficient existing linear depth MCX circuit.

\bigskip
\backmatter

\bmhead{Acknowledgements}

This work was financially supported by the Ministry of Science, Technological Development and Innovation of the Republic of Serbia under contract number: 451-03-65/2024-03/200103.

\bigskip


\begin{appendices}

\section{The unitary operator implemented using the QFT-based MCX}\label{secA}

We will show that the unitary operator implemented using our approach is the same as one implemented by standard MCX. The circuit implements
\begin{equation}
	MCX_n=(I_{2\times 2}\otimes(IQFT_{n_c} P_{+1,n_c} QFT_{n_c}))^\dagger\cdot IQFT_{n} P_{+1,n} QFT_n,\label{opMCX}
\end{equation}
where the number of qubits $n=n_c+1$, and $N=2^n$. Here
\begin{equation}
	(IQFT_{n_c} P_{+1,n_c} QFT_{n_c})^\dagger = QFT_{n_c}^\dagger P_{+1,n_c}^\dagger IQFT_{n_c}^\dagger = IQFT_{n_c} P_{-1,n_c} QFT_{n_c} ,\label{eqhelp}
\end{equation}
where $QFT_{n_c}^\dagger=IQFT_{n_c}$ ($IQFT_{n_c}^\dagger=QFT_{n_c}$), and $P_{+1,n_c}^\dagger=P_{-1,n_c}$.

\begin{figure}[h]
\centering
\includegraphics[width=1\textwidth]{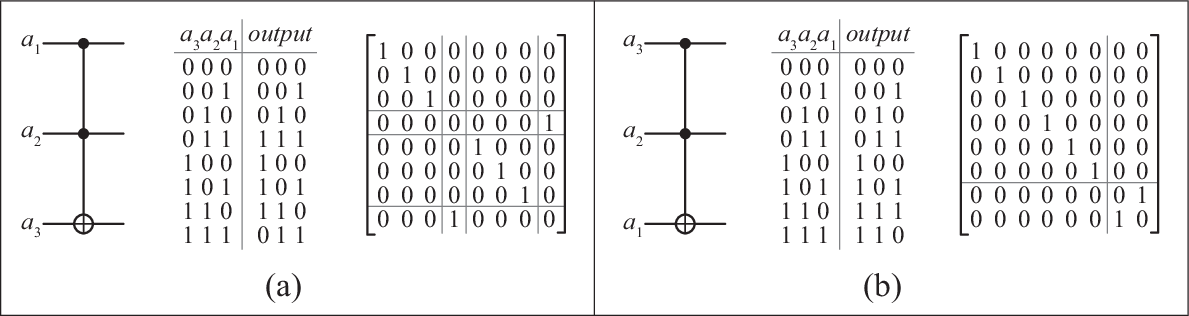}
\caption{The Toffoli gate, truth table, and permutation matrix when the target qubit is (a) the most significant and (b) the least significant.}\label{figA1}
\end{figure}

At first, we manipulated the unitary matrix representation of the QFT, IQFT, and P for $n=3$ and $n=4$ and multiplied matrices according to Eq.~\ref{opMCX} to show it correct in these cases. Using inductive reasoning, results were generalized for an arbitrary $n$. That was very tedious and uneasy to prove. While doing it, we noticed that the result for the $IQFT_{n} P_{+1,n} QFT_n$ was evident:
\begin{equation}
	IQFT_{n} P_{+1,n} QFT_n=
	\begin{bmatrix}
    		0 & 0 & 0 & 0 & 0 & \dots & 0 & 0 & 1 \\
    		1 & 0 & 0 & 0 & 0 & \dots & 0 & 0 & 0 \\
    		0 & 1 & 0 & 0 & 0 & \dots & 0 & 0 & 0 \\
    		0 & 0 & 1 & 0 & 1 & \dots & 0 & 0 & 0 \\
    		\hdotsfor{8} \\
    		0 & 0 & 0 & 0 & 0 & \dots & 0 & 0 & 0 \\
    		0 & 0 & 0 & 0 & 0 & \dots & 1 & 0 & 0 \\
    		0 & 0 & 0 & 0 & 0 & \dots & 0 & 1 & 0
	\end{bmatrix}
	=
	\begin{bmatrix}
		\begin{array}{c|c}
    		\overbrace{0\cdots \cdots \cdots 0}^{N-1} & 1 \\
    		\hline
    		I_{(N-1)\times(N-1)} & 0
    		\end{array}
	\end{bmatrix}, \label{Mp1}
\end{equation}
since it performs $N$-element circular shifts to the left in the computational $Z$-basis.
Similarly,
\begin{equation}
	IQFT_{n_c} P_{-1,n_c} QFT_{n_c}=
	\begin{bmatrix}
		\begin{array}{c|c}
    		0 & I_{\left(\frac{N}{2}-1\right)\times \left(\frac{N}{2}-1\right)} \\
    		\hline
    		1 & \overbrace{0\cdots \cdots \cdots 0}^{\frac{N}{2}-1}
    		\end{array}
	\end{bmatrix} \label{Mm1}
\end{equation}
executes circular shifts to the right. 
Substituting Eqs.~(\ref{Mp1}) and (\ref{Mm1}) into Eq.~(\ref{opMCX}), we obtain
\begin{align}
	MCX_n=&
	\begin{bmatrix}
	\begin{array}{c|c|c|c}
    	I_{\left(\frac{N}{2}-1\right)\times\left(\frac{N}{2}-1\right)} & 0_{\left(\frac{N}{2}-1\right)\times 1} & 0_{\left(\frac{N}{2}-1\right)\times \left(\frac{N}{2}-1\right)} & 0_{\left(\frac{N}{2}-1\right)\times 1} \\
    	\hline
    	0\:0\:0\:0\cdots 0\:0\:0\:0 & 0 & 0\:0\:0\:0\cdots 0\:0\:0\:0 & 1\\
    	\hline
    	0_{\left(\frac{N}{2}-1\right)\times \left(\frac{N}{2}-1\right)} & 0_{\left(\frac{N}{2}-1\right)\times 1} & I_{\left(\frac{N}{2}-1\right)\times\left(\frac{N}{2}-1\right)} & 0_{\left(\frac{N}{2}-1\right)\times 1} \\
    	\hline
    	0\:0\:0\:0\cdots 0\:0\:0\:0 & 1 & 0\:0\:0\:0\cdots 0\:0\:0\:0 & 0
    	\end{array}
	\end{bmatrix}\nonumber\\
	=&C_1C_2\cdots C_{n-2}C_{n-1}-X_n=C^{n_c}-X_n.
\end{align}\label{MCXmatrix}
Using the reversed qubits ordering, one may find familiar notation
\begin{equation}
	MCX_n^{rev}=
	\begin{bmatrix}
		\begin{array}{c|c}
    		I_{(N-2)\times(N-2)} & 0_{(N-2)\times 2} \\
    		\hline
    		0_{2\times (N-2)} & X_{2\times 2}
    		\end{array}
	\end{bmatrix},
\end{equation}
where $X_{2\times 2}$ is the Pauli-X matrix
\begin{equation}
	X_{2\times 2}=
	\begin{bmatrix}
		0 & 1 \\
    		1 & 0
	\end{bmatrix} .
\end{equation}
The gates schematics, truth tables, and permutation matrices for different qubits ordering in the 3-qubit case, which is the Toffoli gate, are given in Fig.~\ref{figA1}.

Based on quantum arithmetics, even one that is not the QFT-based, many complex gates can be very efficiently implemented.

\section{Optimization and decomposition}\label{secB}
\renewcommand\thefigure{B.\arabic{figure}}
\setcounter{figure}{0}
\renewcommand\theequation{B.\arabic{equation}}
\setcounter{equation}{0}

\begin{figure}[h]
\centering
\includegraphics[width=0.95\textwidth]{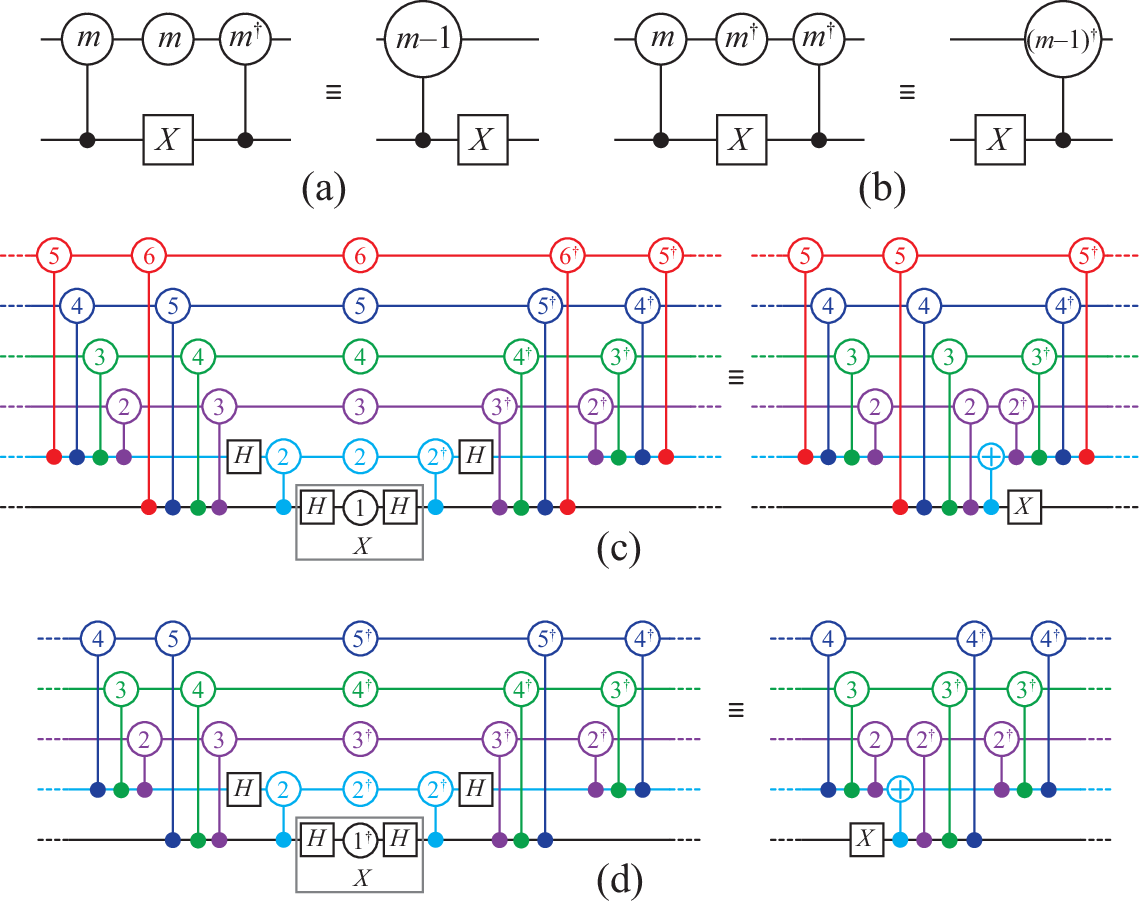}
\caption{The identity used to simplify (a) {\color{RoyalBlue}``$+1$''} and (b) {\color{red}``$-1$''} circuits. Merging of phase gates and controlled phases for 6-qubit MCX in (c) {\color{RoyalBlue}``$+1$''} and (d) {\color{red}``$-1$''} circuits, respectively.}\label{figB1}
\end{figure}

In MCX, there are phase $P_{\pm 1}$ gates between QFT and IQFT. One may note that on the first wireline we have $HR_1H=HZH=X$. Using identities displayed in Figs.~\ref{figB1}(a) and (b) we conclude that we don't need to use any gates to implement $P_{\pm 1}$. Moreover, consecutive controlled phase gates ending/starting QFT/IQFT and phase gates in P will merge. As a result, we will have 4 fewer time slots each in the {\color{RoyalBlue}``$+1$''} and {\color{red}``$-1$''} blocks, as shown in Figs.~\ref{figB1}(c) and (d), respectively.

\begin{figure}[h]
\centering
\includegraphics[width=0.75\textwidth]{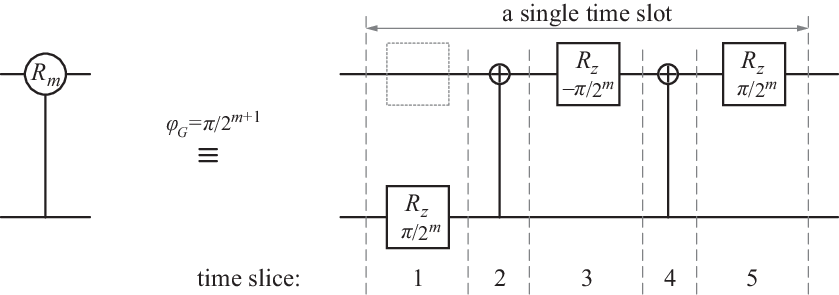}
\caption{The decomposition of $C-R_m$ gate up to the global phase $\varphi_G$ in the IBM's Falcon family processor native gate set.}\label{figB2}
\end{figure}

The native gate set (NGS) used by superconducting IMB's Falcon processor family is $\{C-X, R_z, ID, SX=\sqrt{X}, X\}$. One may show that $R_m=\exp(i\frac{\pi}{2^m})\cdot R_z(\frac{\pi}{2^{m-1}})$, $H=\exp(i\frac{\pi}{4})\cdot R_z(\frac{\pi}{2})\sqrt{X} R_z(\frac{\pi}{2})$ and $SWAP_{12}=(C_1-X_2)(C_2-X_1)(C_1-X_2)$\cite{Nielsen2010}. Due to optimization, we don't use single-qubit $R_m$ gates. Hadamard and SWAP comprise three native gates and use three elementary time intervals to execute. However, we need three $R_z$ and two $C-X$ gates to implement $C-R_m$ \cite{Barenco1995}. It is executed in 5 time sequences, as explicitly shown in Fig.~\ref{figB2}.

We will decompose our circuit into the NGS. This procedure is called {\it transpilation}. Using the transpiled MCX circuit one may find circuit depth (time for circuit execution) and the number of elementary gates used in genuine quantum computation. 
The circuit transpilation process involves some optimizations. By merging consecutive elementary gates of the same type or parallelizing the execution of ones acting on different qubits, we can reduce the experimental circuit's time and space complexity. A single-qubit gate can shift between time slices along the wireline up to the slice with different types of single-qubit gates or up to a $C-X$ gate acting on that qubit. In each QFT there are $C_j-R_{j',m}\cdot C_{j+1}-R_{j',m-1}$ and $C_{j-1}-R_{j,2}\cdot H_j$ subcircuits. It is easy to notice that $R_z(\pi/2^{m-1})$ on the ${j^\prime}^{\rm th}$ wireline  of $C_{j+1}-R_{j',m-1}$ can be executed simultaneously with $R_z(\pi/2^{m})$ on the $j^{\rm th}$ control line of $C_j-R_{j',m}$ gate. The position in the time slice where $R_z$, located in the fifth time slice of the neighboring gate, can ``shift'' is indicated by a dotted line in Fig.~\ref{figB2}. Similarly, $R_z(\pi/2)$ from $H_j$ can be executed simultaneously with $R_z(\pi/4)$ on the control line of $C_{j-1}-R_{j,2}$ (moving to the position in $C-R_m$ denoted by a dotted line in Fig.~\ref{figB2}). This rule mirrored applies to IQFT, too. Also, there are $C_{j+1}-R_{j',m}^\dagger \cdot C_{j}-R_{j',m}$ gates between QFT and IQFT. Here, on the ${j^\prime}^{\rm th}$ controlled line $R_z(\pi/2^m)$ and $R_z(\pi/2^m)^\dagger=R_z(-\pi/2^m)$ annihilate. Using simultaneous executions and gates annihilation, the effective depth of the $C-R_m$ and $H$ gates is reduced to 4 and 2, respectively.

\end{appendices}

\bigskip
\input MCX_QFT_arxiv.bbl



\end{document}